\documentstyle[preprint,aps]{revtex}

        \newcommand{\leftdraft}[1]{
        }

\begin{document}
\vskip 1cm
\leftdraft{hep-ph/9505424}

\title{\hskip 10 cm UAB-FT-367\newline\newline Stability of Z-strings in
  strong magnetic fields} \author
{Jaume Garriga and Xavi Montes} \address{IFAE, Edifici C, Universitat
  Aut\`onoma de Barcelona.E-08193 Bellaterra, Spain.}
\maketitle
\begin{abstract}
  We show that the Z-strings of the standard electroweak theory can be
  stabilized by strong external magnetic fields, provided that $\beta^{1/2}
  \equiv M_H/M_Z\leq 1$, where $M_H$ and $M_Z$ are the Higgs and Z masses.
  The magnetic fields needed are larger than $\beta^{1/2} B_c$ and smaller
  than $B_c$, where $B_c\equiv M^2_W/e$ is the critical magnetic field which
  causes W-condensation in the usual broken phase vacuum.  If such magnetic
  fields were present after the electroweak transition, they would stabilize
  strings for a period comparable to the inverse Hubble rate at that time.
  Pair creation of monopoles and antimonopoles linked by segments of string is
  briefly considered.
\end{abstract}

$Z$-strings are Nielsen-Olesen vortex solutions `embedded' in the electroweak
theory \cite{EWTanmay}. They can be described as flux tubes of the Z gauge
boson trapped in a thin core where the Higgs field vanishes. The Higgs
develops an expectation value outside the core, where its phase has a net
winding around the string.  Like the sphaleron \cite{SphaManton}, Z-strings
can carry baryon number and may have played a role in the generation of baryon
asymetry \cite{BarAsym}. Nambu's early work \cite{EWNambu} showed that a
finite segment of string would be bounded by a magnetic monopole on one end
and an antimonopole at the other.  Detailed studies of the vortex, by
Vachaspati and collaborators \cite{EWTanmay,EWMargaret} stirred a new wave of
attention.  Unlike the vortices in the Abelian Higgs model, which are
topologically stable, $Z$-strings are embedded in a larger theory and their
stability depends on the dynamical details. Unfortunately, the analysis showed
that, for $\sin^2\theta_W\approx .23$, $Z$-strings are unstable
\cite{EWMargaret,Cinderella,HindEW}.  Then, even if they formed at the phase
transition, their lifetime would probably be too short for them to play a
cosmological role.

The above studies, however, ignored the effect of a possible background
magnetic field.  Right after the electroweak phase transition one may expect
strong magnetic fields on small scales \cite{PrimMagn}, and it
is natural to ask whether these would affect the stability of vortices.The aim
of this letter is to show that there is a range for the magnetic field for
which $Z$-strings would be stable.  Intuitively, this result can be understood
as follows.  Recall that the instability of strings is related to the
phenomenon of W-condensation \cite{OnEWMagn,EWStrPerkins}. The strong Z-flux
in the core of the string is coupled to the anomalous magnetic moment of the
W-bosons, and this causes the W's to condensate. We shall see that by opposing
the Z-flux with an external magnetic field, the condensation can be avoided.

To be definite, let us consider the bosonic part of the electroweak energy
functional. For static configurations
\begin{eqnarray}
  \lefteqn{E=\int d^2x\,\left\{ -({D_j}\Phi)^\dagger {D^j}\Phi +
      \lambda\left(|\Phi|^2 - \frac{\eta^2}{2}\right)^2 \right.}\nonumber \\&
  &+\frac{1}{2}{\bf B}_Z^2 +\frac{1}{2}{\bf B}_A^2+ \frac{1}{2}{
    G_{ij}}^\dagger G^{ij} - i g{\bf B}_T \cdot {\bf W}^\dagger \times {\bf W}
  \nonumber\\& &\left. + 2 g^2({\bf W}^\dagger\cdot {\bf W})^2-2 g^2({\bf
      W}^\dagger\cdot{\bf W}^{\dagger})({\bf W}\cdot{\bf
      W})\frac{}{}\right\}.\label{energy}
\end{eqnarray}

Here ${\bf B}_A=\nabla\times{\bf A}$, ${\bf B}_Z = \nabla\times{\bf Z}$,
$g{\bf B}_T = e {\bf B}_A + \gamma{\bf B}_Z$, and $G_{ij}=\partial_{[i} W_{j]}
- i\, e W_{[i}A_{j]}-i\, \gamma W_{[i}Z_{j]}$.  As usual, $g=\alpha
\cos\theta_W$, $e=g\sin\theta_W$ and $\gamma=g\cos\theta_W$, and we follow the
conventions of Ref.~\cite{GaugeThChLi}.

Ambjorn and Olesen have shown \cite{OnEWMagn} that the anomalous mass
term
\begin{equation}
  -i g{\bf B}_T \cdot {\bf W}^{\dagger}\times{\bf W} \label{1}
\end{equation}
acts like a negative contribution to the $W$ mass squared. An instability
develops when the combined field strength $gB_T=|e {\bf B}_A + \gamma
{\bf B}_Z|$ exceeds a certain threshold.  For instance, in the usual vacuum
(without strings), this happens when the magnetic field contribution
exceeds the usual positive $W$ mass term $M_W=g\eta/2$,
$$
e B_A > e B_c \equiv M_W^2. \label{wcondensate}
$$
Perkins, on the other hand, has pointed out
\cite{EWStrPerkins} that in the core of a string the field strength
${\bf B}_Z$
causes W-condensation for the physical value of $\cos \theta_W$.

It is clear that a constant external magnetic field ${\bf B}_A$ antiparallel
to ${\bf B}_Z$ would reduce the anomalous term~(\ref{1}) in the core of the
string, and hence strings might be stabilized in the physical region of
parameter space. Of course, detailed analysis is required before we jump to
conclusions. If we need magnetic fields larger than $B_c$ to counteract $B_Z$
in the core of the string, these would cause $W$-condensation outside the
core.  Also we have to consider all possible modes of fluctuation, not just
the formation of $W$-condensates.

\vskip .2 truecm
{\em Stability analysis}
\vskip .2 truecm

The $Z$-string solution is given by
\begin{equation}
\begin{array}{cc}
\Phi=f(r)e^{in\theta}
\left(\begin{array}{c} 0\\ 1\end{array}\right), & Z=-v(r)d\theta
\end{array},\label{vortex}
\end{equation}
with all the other fields set to zero. The profiles $f(r)$ and $v(r)$
satisfy the Nielsen-Olesen differential equations \cite{StringNO}.
We can always add to the solution a constant magnetic field $B_A$,
with potential
\begin{equation}
A={B_A\over 2}r^2 d\theta.\label{camp}
\label{form}
\end{equation}
This satisfies the field equations without altering
$f(r)$ and $v(r)$, since $\Phi$ and $Z$ are not
magnetically charged. On the other hand,
the fluctuations of $W$ and of the upper
component of the Higgs field are charged and will feel the presence of
$B_A$.

The stability analysis can be carried out as in the case with no magnetic
field \cite{EWMargaret}. Full details will be reported
elsewhere~\cite{XaviTesina}. To check whether the string is a local minimum
of (\ref{energy}), one must consider generic perturbations to
(\ref{vortex}).  It can be shown that the energy is minimized when the
perturbations to the $Z$, $A$, and to the lower component of the
Higgs field are set to zero. The reason can be traced to the fact that the
abelian Nielsen-Olesen string is stable. The perturbed configuration is then
written as
\begin{equation}
\Phi=\left(\begin{array}{c} \phi_1 \\ f(r)e^{in\theta}
\end{array}\right); \quad
W= W_{\mu} dx^{\mu},\label{vortpert}
\end{equation}
with the $Z$ and $A$ fields as given in (\ref{vortex}) and (\ref{camp}). The
perturbations can be taken to be independent of the $z$ coordinate, and the
$z$ components of the gauge fields can be set to zero, since relaxing these
conditions always results in an increase of energy \cite{EWMargaret}. For the
same reasons we have set the temporal components of the gauge fields to zero
in (\ref{energy}). Generic perturbations can then be expanded in Fourier
series
\begin{eqnarray*}
  &\phi_1={\displaystyle \sum_m e^{im\theta}\chi_m(r)},&\\ &W={\displaystyle
    \frac{1}{\sqrt 2}\sum_l e^{il\theta} [\Theta_l (r) d\theta + i R_l (r)
    dr]}&,
\end{eqnarray*}
where $\chi, \Theta$ and $R$ are arbitrary complex functions of $r$.
Substituting into~(\ref{energy}), expanding to quadratic order in the
fluctuations and integrating the angular dependence we find that the energy
fluctuation is $\delta E=\sum_l E_l$. Here $E_l$ is a quadratic functional
depending on $\chi_{n-l}$, $\Theta_l$ and $R_l$. Actually $R_l$ appears only
as a Lagrange multiplier. The energy is minimized by imposing $\delta
E_l/\delta R_l=0$, from which $R_l$ can be eliminated. The functional $E_l$
depends now on two functions $\chi_{n-l}$ and $\Theta_l$, but by diagonalizing
the
kinetic term we readily find that there is only one dynamical degree of
freedom.  Defining the new fields
\begin{eqnarray*}
 &\zeta_l=M\,\chi_{n-l} + (gf/2)\,\Theta_l &\\&\zeta^{(g)}_l=
gfr^2\,\chi_{n-l} - M \,\Theta_l&,
\end{eqnarray*}
where $ M \equiv l + e\,(B_A/2)r^2 - \gamma v$ and $N \equiv M^2+ (gr^2f^2/2)
$, one finds that $\zeta^{(g)}_l$ drops out. This simplification is due to
gauge invariance: it can be shown that $\zeta^{(g)}_l$ corresponds to a pure
gauge mode \cite{XaviTesina,EWMargaret}.  Finally, after lengthy
algebra~\cite{XaviTesina}
$$
E_l\ge  2\pi \int_0^\infty r\, dr\, \zeta_l {\cal O}\zeta_l,
$$
where
\begin{displaymath}
  {\cal O}\equiv-\frac{1}{r}\frac{d}{dr}\left(\frac{r}{N}\frac{d}{dr}\right)
  + {\cal U}(r),
\end{displaymath}
\begin{eqnarray*}
  {\cal U}(r)\equiv\frac{f'^2}{Nf^2}+\frac{2 S}{g^2 r^2f^2} +
  \frac{1}{r}\frac{d}{dr}\left(\frac{rf'}{Nf}\right)\\
  S \equiv \frac{g^2f^2}{2}-\frac{M'^2}{N}-r\frac{d}{dr}
  \left(\frac{M'}{r}\frac{M}{N}\right),
\end{eqnarray*}
and a prime indicates $d/dr$.

Just as in the case with no magnetic fields, the problem reduces to finding
whether the equation ${\cal O} \zeta_l = w \zeta_l$ has any negative
eigenvalues $w$. The boundary conditions of the eigenvalue equation are
$\zeta(r\to 0) \to r^{|n-l|} $ if $l \not=0$ and $\zeta(r\to \infty) \to 0$
for all $l$. For $n=1$ and 2, the $l=0$ mode is never
unstable~\cite{XaviTesina}. For $n\ge3$ and $l=0$ the boundary condition at
the origin is $\zeta\to r^n$.

\vskip .2 truecm
{\em Results for n=1}
\vskip .2 truecm

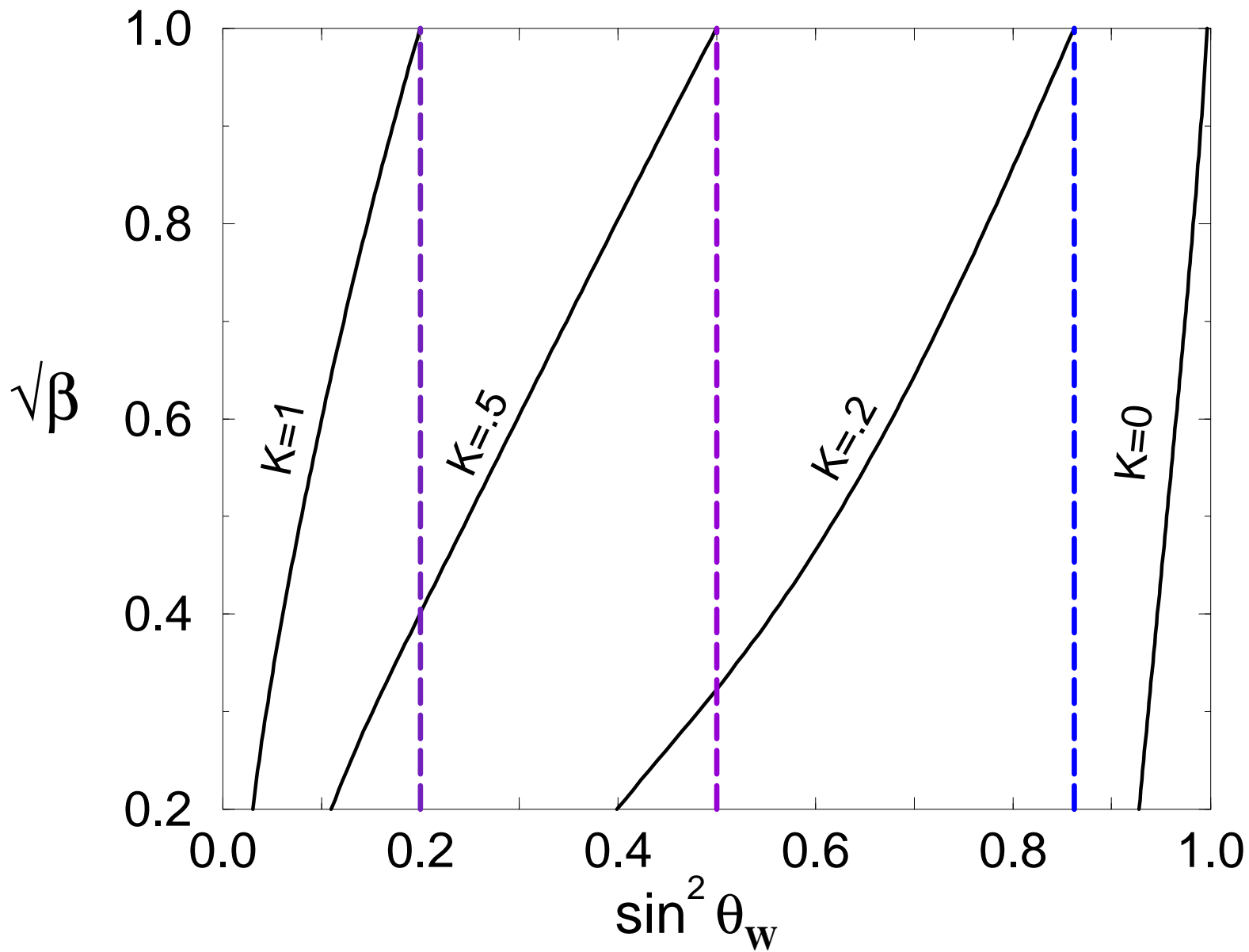
\begin{figure}
\caption{Stability of the $l=1$ mode
    for $n=1$ and different values of the magnetic field $B_A=2\,K
    M_Z^2/\alpha$. The solid curves are critical lines separating the regions
    of stability (right) and instability (left). The vertical lines correspond
    to $B_A=B_c$. For $K\approx 1$, the $l=1$ mode is stable for
    $\sin^2\theta_W\approx .23$}
\label{fig1}
\end{figure}

Fig.~\ref{fig1} summarizes the stability of the l=1 mode for different values
of the magnetic field $B_A=2\,K\,M_Z^2/\alpha$, in the parameter space
$(\beta^{1/2},\sin ^2\theta_W)$.  Here $\beta^{1/2}\equiv M_H/M_Z$, where
$M_H=(2\lambda)^{1/2}\eta$ and $M_Z=\alpha \eta/2$ are the Higgs and Z
masses. The solid curves are the critical lines separating the regions of
stability (right) and instability (left). This instability corresponds to the
formation of a $W$-condensate in the core of the string due to $B_Z$
\cite{EWStrPerkins}.

For $B_A=0$ the result coincides with the critical line found in Ref.
\cite{EWMargaret}.  As the magnetic field $B_A$ is increased, the critical
line moves to the left, so that this mode is stable for smaller values of
$\sin^2\theta_W$. The interpretation is that the magnetic field opposes $B_Z$
and hence $W$-condensation is inhibited. In particular, when $K\approx 1$ the
$l=1$ mode is stable for $\sin^2\theta_W\approx .23\ $.  The vertical lines
correspond to $B_A=B_c$.  To the right of the vertical lines we have
$W$-condensation in the trivial vacuum, outside the core of the string
\cite{OnEWMagn}.  Note that for $\beta=1$ the critical lines $l=1$ meet with
the vertical lines. This is consistent with Ref. \cite{ZFlux}, where the
Bogomol'ny limit $(\beta=1)$ was considered.

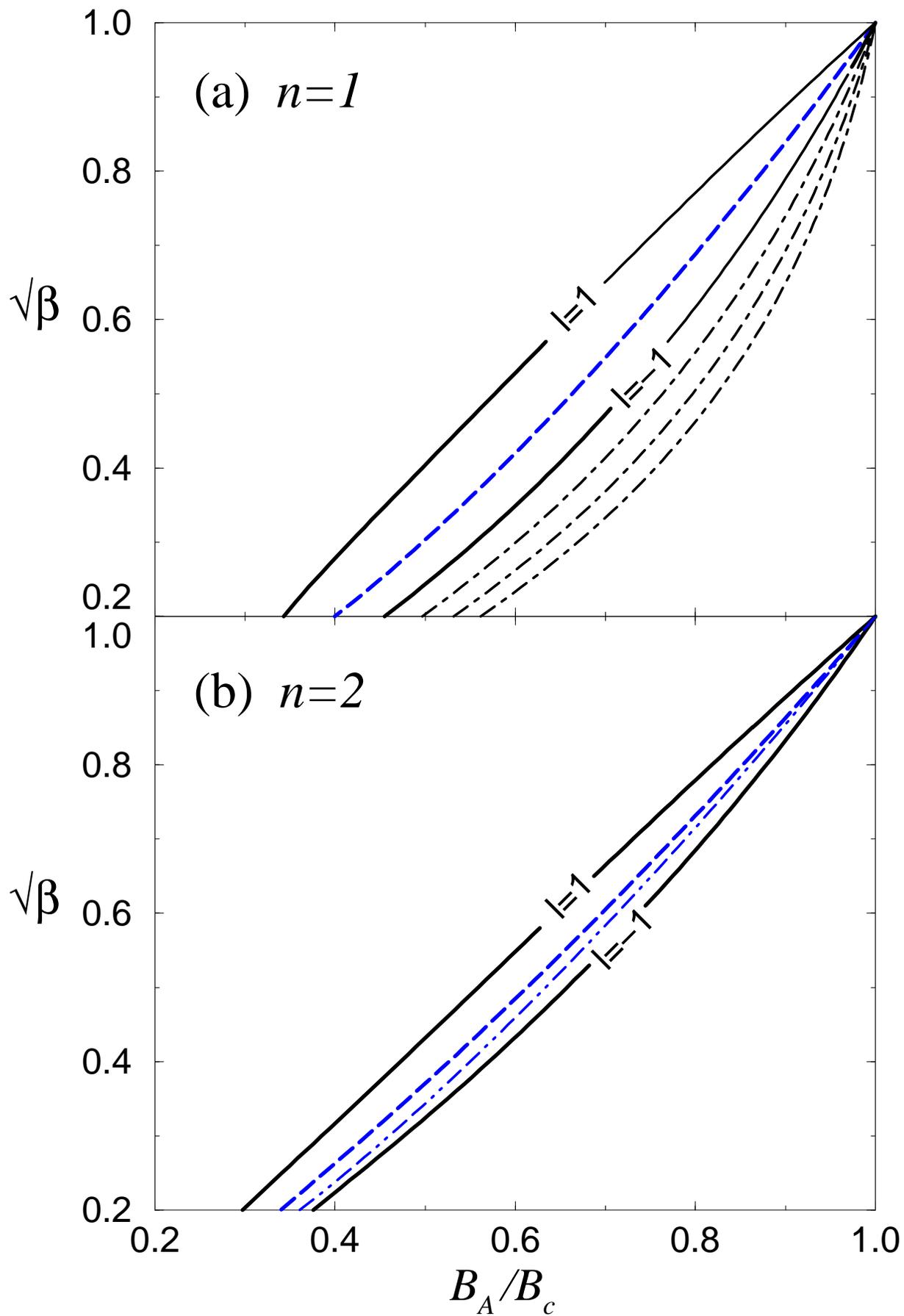
\begin{figure}
\caption{The critical lines of
    stability for $l=1$ and $l=-1$ are shown as solid lines, for $n=1$ and
    $2$, with $\sin^2\theta_W=.23$. Strings are metastable in the region
    between solid lines.  This region is centered around the locus where
    $B_T=0$ at $r=0$, depicted as a dashed line. For $n=1$, the critical lines
    for $l=-2,-3,-4$ are shown in dot-dashed lines. The corresponding
    instabilities develop to the right of these lines, so they do not restrict
    the stable region any further. For $n=2$ the transition line $Q
    B_A=\mu_2-\mu_1$ is depicted as a double dot-dashed line.}
\label{fig2}
\end{figure}

However, this is not the whole story, since we have to consider all values of
$l$. When $eB_A$ is larger than $\gamma B_Z$ in the core of the string,
W-condensation develops in the direction {\em opposite} to the one it did
when $B_A=0$. This corresponds to an instability in the mode $l=-1$. In
Fig.~\ref{fig2}a we plot the critical stability lines for $l=1$ and $l=-1$
(solid lines) as a function of the applied magnetic field, for the physical
value of $sin^2\theta_w$. In the region between both lines the strings are
(meta-)stable. The dashed line corresponds to ${\bf B}_T=0$ at $r=0$.  Near
this line the anomalous mass term almost vanishes inside the core, and hence
it is not surprising that we find stability.

We have checked that for $n=1$ the $l=0$ mode is stable for all values of
$B_A$. Some critical lines for $|l|>1$ are also plotted in dot-dashed lines in
Fig. \ref{fig1}. They do not restrict the stability region any further.

The experimental lower bound on the Higgs mass gives
$\beta^{1/2}{\buildrel >\over \sim} .6$. For $.6 {\buildrel <\over
\sim} \beta^{1/2} \leq 1$ the $n=1$ strings are stable over a range of
magnetic fields of relative width
\begin{equation}
{\Delta B_A\over B_A}\approx .2 (\beta^{-1/2}-1).
\label{range}
\end{equation}
If the magnetic field is frozen in with the plasma, as it
is usually assumed \cite{PrimMagn}, we have $B_A\propto a^{-2}(t)$,
where $a(t)\propto
t^{1/2}$ is the cosmological scale factor. Given a sufficiently large
initial magnetic field, this means that strings can be metastable for a
time of cosmological order:
$
\Delta t\approx (\Delta B_A/ B_A) t \approx .2 (\beta^{-1/2}-1)t.
$
This may be sufficient for strings to fall out of thermal equilibrium,
which in turn may have consequences for baryogenesis \cite{BarAsym}.

For $\beta^{1/2}>1$ and $B_A<B_c$ there are no stable Z-strings. When
$B_A>B_c$ the vacuum decays into a lattice of W-condensate vortices
\cite{OnEWMagn}. Since the W bosons couple to the Z flux, we expect that the
simple Z-string solution will not exist in this background.

\vskip .2 truecm
{\em Results for higher winding}
\vskip .2 truecm

Fig.~\ref{fig2}b shows the metastable region for a string of winding number
$n=2$. The critical lines for $l=1$ and $-1$ are shown as solid lines. The
$l=0$ mode is stable in the whole range, and for $|l|>1$ the critical lines
are less restricitive than the ones depicted. The metastable region, between
solid lines, is appreciably narrower than in the case $n=1$. Like in that
case, the region is centered around the locus where ${\bf B}_T=0$ at $r=0$
(dashed line). Of course the line is not the same as in Fig.~\ref{fig2}a,
because $B_Z$ is different.

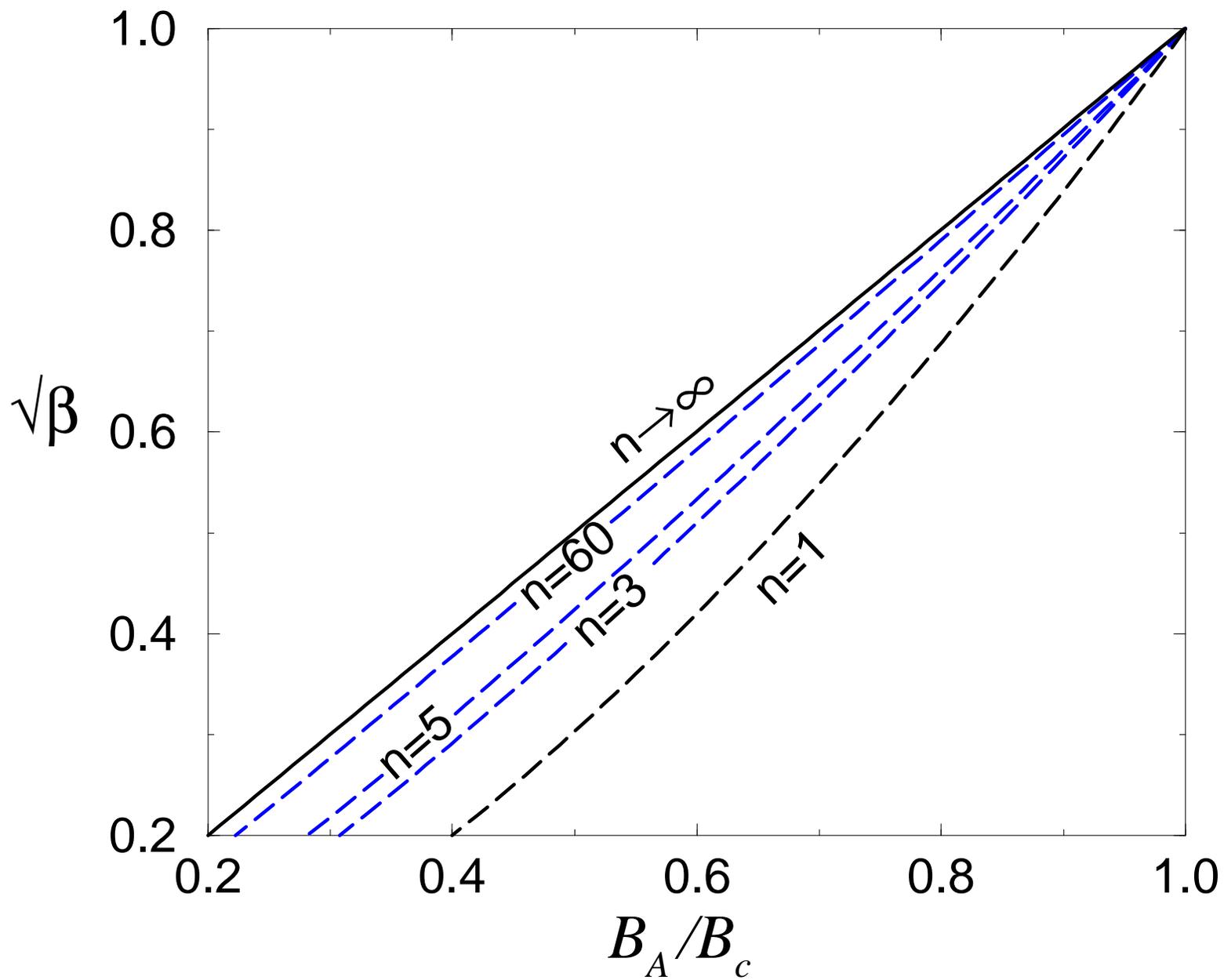
\begin{figure}
\caption{Locus where $B_T=0$ at $r=0$ for some windings. For each $n$ the
  stability region is in the vicinity of these lines.}
\label{fig3}
\end{figure}

The stable regions for $n=1$ and $n=2$ have some overlap, but
the case $n=2$ extends to somewhat lower values of the applied magnetic field.
This ``overlap'' pattern continues for higher $n$. The loci where the
anomalous term vanishes at $r=0$ are represented in Fig.~\ref{fig3} for
different values of $n$. As $n$ increases, the dashed line moves to
lower values of the magnetic field. The stability regions for each $n$
are in the vicinity of these lines. We find \cite{XaviTesina} that the width
of these regions decreases with $n$, but the stable patch for
winding $n$ always has some overlap with the stable patch for
winding $n+1$, which in turn extends to somewhat smaller values of the
magnetic field. For $n\to \infty$ the dashed line tends to the straight
line $B_A=\beta^{1/2}B_c$, and the width of the stable region tends to
zero. Hence $B_A=\beta^{1/2}B_c$ is the lowest value of the magnetic
field for which we can find metastable strings of any winding.

As we cross the critical lines $l=\pm 1$ into the unstable region, the strings
may unwind and release their energy in the form of linear excitations on the
broken phase vaccuum, or settle down into a possible ``dressed string''
configuration\cite{Dressed,Cinderella}. The pattern of stability described
above may also suggest that, as magnetic field decreases, the $n=1$ string
becomes unstable and decays into a metastable $n=2$ string. By decreasing
$B_A$ further, this decays into a $n=3$ string, and so forth. In principle, it
is not clear whether this cascade is energetically allowed; $n=2$ strings are
heavier than $n=1$ strings, but we have to take into account that these
strings end in monopoles of twice the charge (in other words, energy can be
extracted from the magnetic field). We shall return to this question below.

\vskip .2 truecm
{\em Pair creation}
\vskip .2 truecm

The magnetic fields needed above have an energy density comparable to the
electroweak scale, and one  should worry about instabilities of the broken
phase in the presence of such fields. As we shall see, for
$\beta^{1/2}B_c<B_A<B_c$ the broken phase is only metastable.

It is well known that a magnetic field can pair-produce monopoles and
antimonopoles \cite{afma}. Electroweak monopoles are linked by
Z-strings, so pair production can only happen if the force exerted by
the magnetic field on the monopoles exceeds the string tension
\cite{previ},
\begin{equation}
n QB_A > \mu_n. \label{transition}
\end{equation}
Here $Q=(4\pi/\alpha)\tan \theta_W$ is the $n=1$ monopole charge
\cite{EWNambu}, and $\mu_n$ is the string tension. For large $n$ the tension
is dominated by the constant energy density in the core of the string, and we
can neglect gradient terms
$$
\mu_n=S_n\left({\lambda\eta^4\over 4}+{B_Z^2\over 2}\right).
$$
Here $S_n$ is the area of a cross section of the core of the string, and
$B_Z=\phi_n/S_n$, where $\phi_n=4\pi n/\alpha$ is the quantized flux.
Minimizing the tension with respect to $S_n$ we find
$\mu_n=\beta^{1/2}\pi\eta^2 n$. Comparing to the magnetic force, we conclude
that in the large $n$ limit the strings will grow provided that
$B_A>\beta^{1/2} B_c$ (larger values of $B_A$ are needed for lower $n$).

The decay of the metastable phase can proceed through pair production by
thermal activation or by quantum tunneling \cite{afma,previ}.  However, the
actual decay rate and the ``fate'' of this ``false vacuum'' are difficult to
assess.  The problem is that condition~(\ref{transition}) is only satisfied in
the regions where the Z-strings of winding $n$ are unstable. As a result, the
corresponding instantons will have too many negative modes, and will probably
not represent a semiclassical channel of decay \cite{coleman}.  Further work
will be needed to clarify this issue.

Finally, we may ask whether a $n=1$ string can decay into a $n=2$ string.
This process can be thought of as the nucleation of a monopole antimonopole
pair on top of an existing $n=1$ string.  For this to be energetically
possible, we need $QB_A>\mu_2-\mu_1$.  The transition line $QB_A=\mu_2-\mu_1$
is depicted in Fig.~\ref{fig2}b as a double dot-dashed line. It happens to lie
in the intersection of the regions of metastability for $n=1$ and $n=2$. Since
the strings involved are both metastable, this process may be mediated by and
instanton with just one negative mode, as is required by the general theory of
semiclassical tunneling \cite{coleman}.

\vskip .3 truecm

We are grateful to Ana Ach\'ucarro and Tanmay Vachaspati for stimulating
discussions.

%%  Figure Captions

\end{document}